\documentclass[preprint,5p,times,twocolumn]{elsarticle}

\usepackage{amssymb}
 \usepackage{amsthm}

\journal{Physics Letters A}

\begin{document}

\begin{frontmatter}



\title{Exact solutions of the Grad-Shafranov equation via similarity reduction and applications to magnetically confined plasmas}

\author{Dimitrios A. Kaltsas}

\ead{dkaltsas@cc.uoi.gr}
\author{George N. Throumoulopoulos}
 \ead{gthroum@cc.uoi.gr}

\address{%
Department of Physics, University of Ioannina\\
GR 451 10 Ioannina, Greece
}%

\date{\today}
\begin{abstract}
 We derive exact solutions of a linear form of the Grad–Shafranov (GS) equation, including incompressible
equilibrium flow, using ansatz-based similarity reduction methods. The linearity of the equilibrium
equation allows linear combinations of solutions in order to obtain axisymmetric MHD equilibria with
closed and nested magnetic surfaces which are favorable for the effective confinement of laboratory
plasmas. In addition, employing the same reduction methods we obtain analytical solutions for several
non-linear forms of the GS equation. In this context analytic force-free solutions in both linear and
nonlinear regimes are also derived.
\end{abstract}

\end{frontmatter}

\section{The Grad-Shafranov equation}
The GS equation governs the axisymmetric MHD equilibria of static plasmas.  Essentially it is a second order elliptic, generally nonlinear PDE, whose solutions reveal the topology of the magnetic field on the poloidal cross section of a toroidal plasma. In the case of incompressible flows, parallel to the magnetic field, the equation remains identical in form \cite{Throum_gen_GS,Sim}. Also, in \cite{Throum_gen_GS} and \cite{Sim} the equation was further generalized  for incompressible  flows of arbitrary direction with an additional $r^4$-term associated with the equilibrium electric field (see also Eq. (\ref{gen_GS}) below). In cylindrical coordinates $(r,\phi,z)$ the GS equation takes the familiar form
\begin{equation}
\partial_{rr}u-r^{-1}\partial_r u +\partial_{zz}u+f(u)+g(u)r^2=0 \label{GS}
\end{equation}
with $f(u)$, $g(u)$ being free-arbitrary functions which respectively are related to the poloidal current and the plasma pressure. Specifically $g(u)=\mu_{0}P_s'(u)$ where $P_s$ is the pressure in the absence of macroscopic flows. In the presence of field aligned flows the total plasma pressure is given by the Bernoulli equation $P=P_s(u)-\rho v^2/2$, where $v$ is the magnitude of flow velocity and $\rho$ the mass density. The dependent variable $u$ is related to the poloidal flux function $\psi$ through the transformation 
\begin{equation}
u(\psi)=\int_0^{\psi}\left[1-M_p^2(s)\right]^{1/2}ds
\end{equation}
where $M_p(s)$ is the poloidal Mach function \cite{Sim}. Once the free functions $f(u)$ and $g(u)$ are specified the solutions of (\ref{GS}) provide us all the information about the macroscopic equilibrium properties of the magnetofluid. 

So far analytical solutions of (\ref{GS}) have been constructed and studied extensively for choices of constant $f(u)$ and $g(u)$. The corresponding solutions describe the so-called Solovev equilibrium, which is characterized by monotonic current density and safety factor profiles. Solutions for free functions proportional to $u$, expressed in terms of Coulomb or Whittaker wave functions, have been obtained for the first time by Hernneger \cite{Hern} and independently by Maschke \cite{Mas}. The work of Atanasiu et al \cite{Atanasiu} was focused on the case of free functions linear in $u$, computing the required particular solutions of the inhomogeneous PDE with a series-expansion-based technique. Wang \cite{Wang} employing also linear ansatzes, reduced the GS equation into a Helmholtz equation and derived analytical solutions describing equilibria with toroidal current reversals (TCR). In addition, particularly limited progress has been made in the analytical integration of (\ref{GS}) for nonlinear free functions; in the works of Cicogna and Pegoraro \cite{Cic_Peg_1,Cic_Peg_2} classical and non-classical symmetries of special nonlinear forms of the GS equation were exploited for the derivation of analytical solutions. These solutions were implemented on the construction of nonlinear equilibria.

In this letter, exploiting the translational symmetry of the GS
equation, we convert it into a system of two first order PDEs.
This system contains all solutions of the GS equation and thus the
original problem is reduced to the solution of the two first order
PDEs. This is a quite interesting procedure since in several cases it might be more convenient to seek solutions or reduction ansatzes
of the reduced equations rather than of the original PDE. In our
case it wasn't difficult to find a simple ansatz to further reduce
this system. We show that the original PDE results in a single ODE
in terms of a reduction variable which depends on the choice of
the aforementioned ansatz. This procedure serves as a prototype
for algorithmic derivation of reductions, that can possibly be exploited employing additional symmetries which are admitted for
some particular choices of the free functions. Regarding the derived solutions, in the linear case the principle of superposition
allows a linear combination of the derived solutions which results in magnetic configurations with closed and nested magnetic
surfaces, essential for the confinement in fusion devices. However
this is not the case for the corresponding nonlinear solutions and
therefore the magnetic field configurations do not form closed surfaces.

\section{Similarity reduction procedure} 
A similarity reduction is a procedure by which one may find a
``grouping" of the independent variables so as to assume a functional form of the solution that enables the original PDE to be
reduced to a simpler form, usually to an ODE. Our treatment which
finds such a reduction form for the Grad–Shafranov equation was
initially inspired by the work of Anco et al \cite{sigma} focused on the
reduction of a certain class of semilinear reaction–diffusion equations by a method known as ``group foliation". In general a group
foliation converts a PDE into an equivalent system of lower order PDEs, called group resolving system, whose independent and
dependent variables correspond respectively to the invariants and
the differential invariants of a given infinitesimal symmetry generator. In other words a group of point symmetry transformations
induces a foliation of the solution space. Each solution of the group
resolving system corresponds to a one-parameter family of exact
solutions of the original PDE. The interested reader may refer to \cite{sigma} and \cite{ovs}.

In our case due to the arbitrariness of the free functions $f(u)$ and $g(u)$ the only symmetry which is admitted in general by the GS equation is the z-translational symmetry. We show that exploiting this symmetry and following the treatment described in \cite{sigma} we can reduce Eq. (\ref{GS}) into a system of first order, coupled PDEs whose independent and dependent variables are the invariants and the differential invariants of the z-symmetry infinitesimal generator $X_t=\partial_z$. The resulting PDEs can be split into a system of ODEs using suitable separation ansatzes. Under the assumption $g(u)=\epsilon f(u)$ this method yields similarity reductions of the GS equation. The invariants $I$ of $X_t$ satisfy $X_t(I)=\partial_z I=0$. In the space of independent and dependent variables $(r,z,u)$ we can identify two such invariants namely $\gamma=r$ and $\vartheta=u$. The differential invariants $\tilde{I}$ of the z-symmetry infinitesimal generator are quantities  invariant under the action of $X_t$ on the extended, by the first order derivatives of $u$, space that is $(r,z,u,u_r,u_z)$. We are making use of a mixed notation for the partial derivatives so as to underline the fact that $u_r:=\partial_ru$ and $u_z:=\partial_zu$ serve also as new variables. The action of $X_t$ on the extended space is realized through its first order prolongation $pr^{(1)}X_t$ which for the translation symmetry is $X_t$ itself i.e. $pr^{(1)}X_t=\partial_z$. For a detailed description one may refer to \cite{ovs} and \cite{olver}. Therefore the additional invariants provided by the action of $X_t$ on the extended space are $\Gamma:= u_r$, $\Theta:=u_z$.
Obviously the relation $\partial_z u_r=\partial_r u_z$ must hold and
the invariantized variables should also satisfy Eq. (\ref{GS}). Hence we obtain the following system of two coupled PDEs for the functions $\Gamma(\gamma,\vartheta)$ and $\Theta(\gamma,\vartheta)$
\begin{eqnarray}
\partial_{\gamma}\Theta+\Gamma\partial_{\vartheta}\Theta-\Theta\partial_{\vartheta}\Gamma=0 \nonumber \\
\partial_{\gamma}\Gamma+\Gamma\partial_{\vartheta}\Gamma+\Theta\partial_{\vartheta}\Theta-\gamma^{-1}\Gamma+g(\vartheta)\gamma^2+f(\vartheta)=0 \label{grs}
\end{eqnarray}
called translation group resolving system (GRS). The GRS can easily be reduced into a system of ODEs, employing the ansatz $\Gamma=a \gamma Q(\vartheta)$, $\Theta=b W(\vartheta)$ where $a$, $b$ are constants. This choice is based upon the simple observation that if one assumes $\Theta$ as a function of $\vartheta$ only and $\Gamma$ has a separable form then the first equation of the system above becomes automatically an ODE. Also the term $\gamma^{-1} \Gamma$ of the second equation of the GRS becomes a function of $\vartheta$ only if $\Gamma$ is linear in $\gamma$. The resulting ODEs are, 
\begin{eqnarray}
Q'(\vartheta)W(\vartheta)=W'(\vartheta)Q(\vartheta)\nonumber \\
\left(W^2(\vartheta)\right)'=-2b^{-2}f(\vartheta) \nonumber \\
\left(Q^2(\vartheta)\right)'=-2a^{-2}g(\vartheta) \label{ODE_sys}
\end{eqnarray}
The first equation is satisfied for $W(\vartheta)=c_0Q(\vartheta)$, with $c_0=\pm 1$ by choice, which, from the remaining equations, yields $g(\vartheta)=\epsilon f(\vartheta)$ with $\epsilon=a^2/b^2$. Consequently $Q$ is given by 
\begin{equation}
Q(\vartheta)=\pm \left[c-2b^{-2}F(\vartheta)\right]^{1/2} \label{Q}
\end{equation}
where $F$ is the antiderivative of $f$ i.e. $F(\vartheta)=\int_0^\vartheta f(s)ds$ and $c$ is an arbitrary constant representing every constant term into the square root. The solution $u(r,z)$ is then obtained by  integrating the system:
\begin{eqnarray}
\partial_r u(r,z)=\pm ar\left[c-2b^{-2}F\left(u(r,z)\right)\right]^{1/2}\nonumber \\
\partial_z u(r,z)=\pm b \left[c-2b^{-2}F\left(u(r,z)\right)\right]^{1/2} \label{gf_sys}
\end{eqnarray}
The resulting solutions are expressed in terms of an arbitrary integration constant $\tilde{c}$ which reflects the invariance of the solutions under z-translations. From system (\ref{gf_sys}) one deduces that the following equation must hold:
\begin{equation}
b\partial_r u(r,z)=\pm ar\partial_z u(r,z) \label{def_1}
\end{equation}
and therefore the solutions should have the form 
\begin{equation}
u(r,z)=w(x),\qquad with \qquad x:=ar^2/2\pm bz \label{def_2}
\end{equation}
Substituting (\ref{def_2}) into (\ref{gf_sys}) one obtains 
\begin{equation}
w'(x)=\pm\left[c-2b^{-2}F(w(x))\right]^{1/2} \label{w}
\end{equation}
Since $u$ can be written as an arbitrary function of a single variable $x$ one may search directly for solutions of the original PDE in terms of this variable. This is the concept of the direct reduction method \cite{Clar_Krus,Litvi} which in general seeks for solutions of the form $u(r,z)=U\left(r,z;w(x(r,z))\right)$ trying to identify the appropriate forms of the function $U$, and of the reduction variable $x$ so as to reduce the PDE into an ODE for $w(x)$. If we substitute $u=w(x(r,z))$  directly in GS equation we obtain:
\begin{eqnarray}
w'(x)\left(\partial_{rr}x+\partial_{zz}x-r^{-1}\partial_rx\right)+\nonumber \\
w''(x)\left[\left(\partial_rx\right)^2+\left(\partial_zx\right)^2\right]+f(w)+g(w) r^2=0\label{GS_red}
\end{eqnarray}
Substituting $x:=ar^2/2\pm b z$ one can observe that Eq. (\ref{GS_red}) splits into two identical equations if $g(u)=\epsilon f(u)$, where $\epsilon=a^2/b^2$.
The function $w(x)$ is then determined by the ODE:
\begin{equation}
w''(x)+b^{-2}f\left(w(x)\right)=0 \label{dir_red_ode}
\end{equation}
which can easily be transformed to Eq. (\ref{w}) if we express the second order derivative of $w(x)$ as $w''(x)=\frac{1}{2}\frac{d}{dw}\left[\left(w'(x)\right)^2\right]$.
A direct reduction may not always be a convenient approach since in general it is not easy to identify the appropriate reduction variable, particularly when it assumes more complicated forms. Using the method of group foliation we established the framework for a systematic, algorithmic derivation of a reduced representation of the GS equation that is its resolving system, which along with an easily identifiable ansatz led us systematically to (\ref{w}). It is also possible that additional ansatzes may be suitable for reducing the resolving system.
Before closing this section note that from Eq. (\ref{GS_red}) we can also realize that for zero pressure-gradient equilibria, i.e. $g(u)=0$, a reduced ODE occurs if the reduction variable is modified by adding a $az^2$ term, namely $x=a(r^2+z^2)\pm bz$. Then the corresponding version of  Eq. (\ref{dir_red_ode}) is 
\begin{equation}
(4ax+b^2)w''(x)+2aw'(x)+f(w(x))=0 \label{ff_ode}
\end{equation}
\section{Exact solutions}
\subsection{Linear solutions}We first examine the case of the linearized Grad-Shafranov equation with free functions of the form $f(u)=f_1+f_2u$ and $g(u)=g_1+g_2u$ which correspond to poloidal current and the static pressure functions containing both linear and quadratic terms in $u$. The inclusion of quadratic terms is  very interesting since equilibria with hollow toroidal current densities, which can be constructed due to the quadratic terms, are related to the formation of internal transport barriers (ITBs) \cite{conor} which reduce the transport effects and contribute in the transition to high confinement modes. Also this model can describe core-TCR scenarios which thus far have been studied mainly numerically, e.g. \citep{Mart,Rod_Biz_1,Rod_Biz_2} or in the large aspect ratio limit \cite{Martins}. Speaking of TCR equilibria we mention that an alternative method of spectral representation of the flux surfaces has been introduced in \cite{lud}. As we have already mentioned  the methods exposed above are applicable in the case $g(u)=\epsilon f(u)$, hence the general linearized GS equation contains 3 instead of 4 free parameters, 
\begin{equation}
\partial_{rr}u-r^{-1}\partial_r u+\partial_{zz}u+f_1+f_2u+ g_2\left(\frac{f_1}{f_2}+u\right)r^2=0\label{lin_GS}
\end{equation}
where $f_1$, $f_2$ and  $g_2$ are the free parameters and $\epsilon:=\frac{g_2}{f_2}$.
The linear solutions occur by solving Eq. (\ref{w}) or Eq. (\ref{dir_red_ode}) for $f(w)=f_0+f_1w$ and substituting $x=\sqrt{\epsilon} b r^2/2 \pm  bz$,
\begin{eqnarray}
u^{\pm}_1(r,z)=\lambda_1cos\left(\frac{\sqrt{g_2}}{2} r^2\pm\sqrt{f_2}z\right) \nonumber \\
+\lambda_2sin\left(\frac{\sqrt{g_2}}{2} r^2\pm\sqrt{f_2}z\right)-\frac{f_1}{f_2}\label{sol_lin_1}
\end{eqnarray}
 where $\lambda_{1,2}$ are arbitrary constants. This solution generalizes by an inhomogeneous constant term, a solution to the homogeneous counterpart of Eq. (\ref{lin_GS}), derived using a different approach in \cite{Cic_Peg_1}. Due to the ``phase" argument $ar^2/2\pm bz$ each one of the solutions $u_1^{\pm}$ exhibits a parabolic topology on the $(r,z)$-plane. However the linearity of Eq. (\ref{lin_GS}) allows one to take the superposition of the two solutions, i.e.
\begin{equation}
u_1(r,z)=s^+u_1^+(r,z)+s^-u_1^-(r,z) 
\end{equation}
where $s^{\pm}$ are constants satisfying $s^++s^-=1$ and which define the up-down asymmetry of the magnetic surfaces. The resulting solution displays a (r, z)-plane topology in connection with configurations with non nested magnetic surfaces, i.e. having magnetic lobes (or islands) similar to those in Fig. 2 of Ref. \cite{Cic_Peg_1}. In Fig. 1(a)-(b) are depicted two isolated sets of closed and nested magnetic surfaces which can describe equilibria of magnetic confinement systems by choosing appropriately the boundary to coincide with the outer close magnetic surface (see caption of Fig. 1). For $s^+=s^-$ these isolated sets are up-down symmetric with respect to the plane $z=0$. In this linear case we can assume $u(r,z)=u_h(r,z)+w(x(r,z))$, with $u_h$ being an arbitrary nonconstant function of $r$ and $z$. The function $w(x)$ should satisfy the ODE (\ref{dir_red_ode}) with $f(w)=f_0+f_1w$ and $u_h$ the homogeneous counterpart of Eq. (\ref{lin_GS}). Following the treatment of \cite{Evan_Throum} we can write the homogeneous solution as a sum of an arbitrary number of terms,
\begin{eqnarray}
u_h(r,z)=\sum_{j=1}^N\bigl[ \bigl. \kappa_{1,j} W_{\nu_j,\frac{1}{2}}(\varrho)cos(jz)+
 \kappa_{2,j}W_{\nu_j,\frac{1}{2}}(\varrho)sin(jz)\nonumber \\
+\kappa_{3,j} M_{\nu_j,\frac{1}{2}}(\varrho)cos(jz)
 +\kappa_{4,j}M_{\nu_j,\frac{1}{2}}(\varrho)sin(jz)\bigl. \bigr]+c.c. 
\end{eqnarray}
where $W_{\nu_j,\frac{1}{2}}$, $M_{\nu_j,\frac{1}{2}}$ the Whittaker functions with $\nu_j=i\frac{j^2-f_2}{4\sqrt{g_2}}$, $\varrho=i \sqrt{g_2}r^2$; $\kappa_{1,j},...,\kappa_{4,j}$ with $j=1,..,N$, are arbitrary parameters and $i^2=-1$. The limit $N$ depends on the number of boundary-shaping and other equilibrium conditions which one desires to impose. Having determined $u_h$, the complete solution of Eq. (\ref{lin_GS}) is:
\begin{equation}
 u_2=u_h+u_1
\end{equation}
We can also take into account a non-parallel flow  contribution \cite{Throum_gen_GS, Sim} modifying our solutions as follows
\begin{equation}
\tilde{u}_{1,2}(r,z)=u_{1,2}(r,z)-\frac{h_1}{g_2} r^2+\frac{h_1f_2}{g_2^{2}} \label{sol_with_flow}
\end{equation}
where the parameter $h_1$ is related to the radial electric field and the function $\tilde{u}$ is a solution of the linear generalized GS equation: 
\begin{eqnarray}
\partial_{rr}u-r^{-1}\partial_r u+\partial_{zz}u+f_0+f_2u\nonumber \\+g_2\left(f_1f_2^{-1}+u\right)r^2+h_1 r^4=0  \label{gen_GS}
\end{eqnarray}
with $f_0=f_1-h_1f_2^2g_2^{-2}$. The term $h_1r^4$ is due to the non-parallel flow contribution which from the Ohm's law induces a finite electric field.
Using $\tilde{u}_{1}$ one can obtain magnetic field topologies which correspond both to compact and non-compact configurations, while exploiting $\tilde{u}_2$ we can construct equilibria with shaped boundaries. These three posibilities are discussed below.
\begin{figure}[h]
\begin{center}
\includegraphics[scale=0.345]{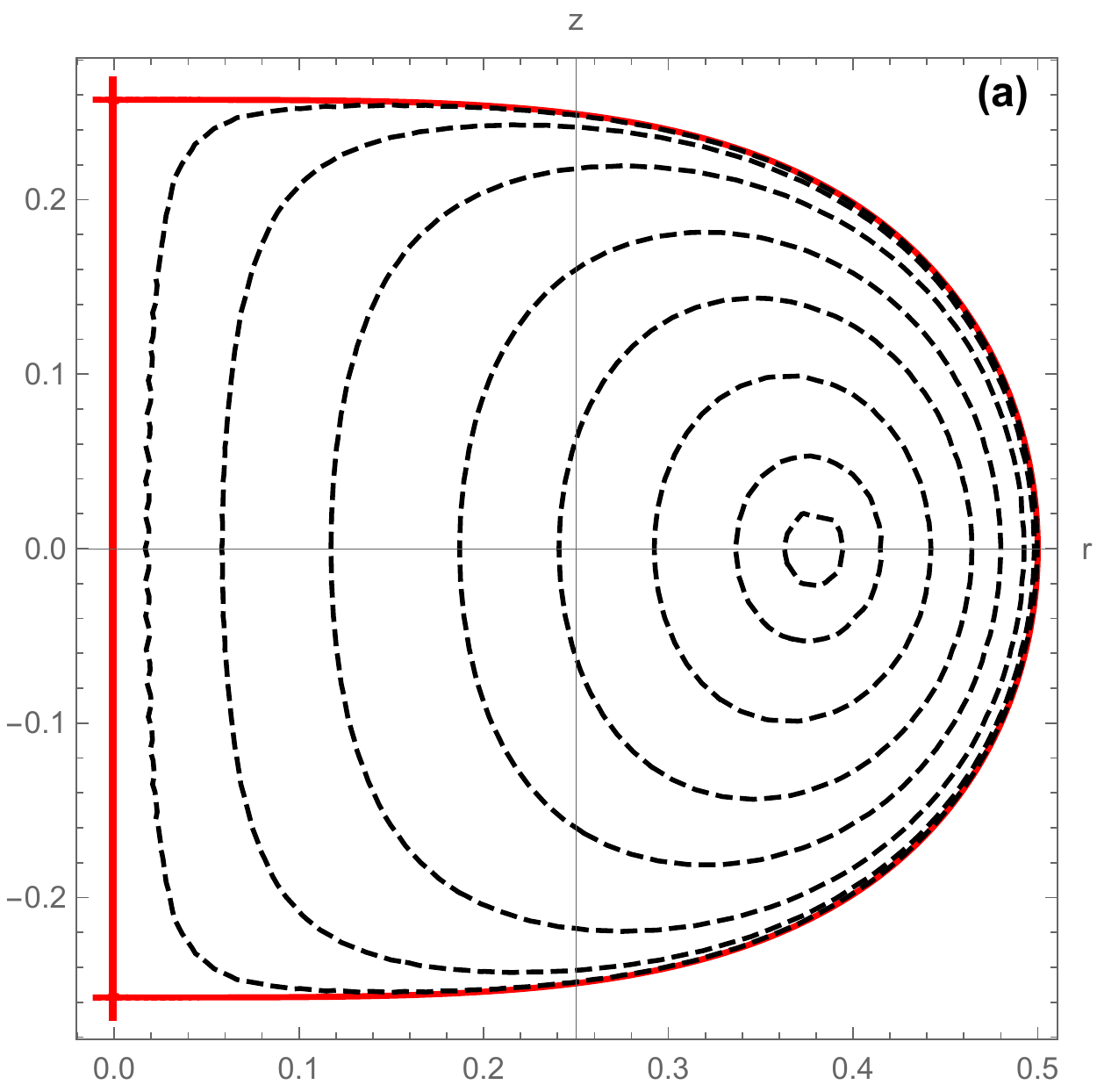} \includegraphics[scale=0.345]{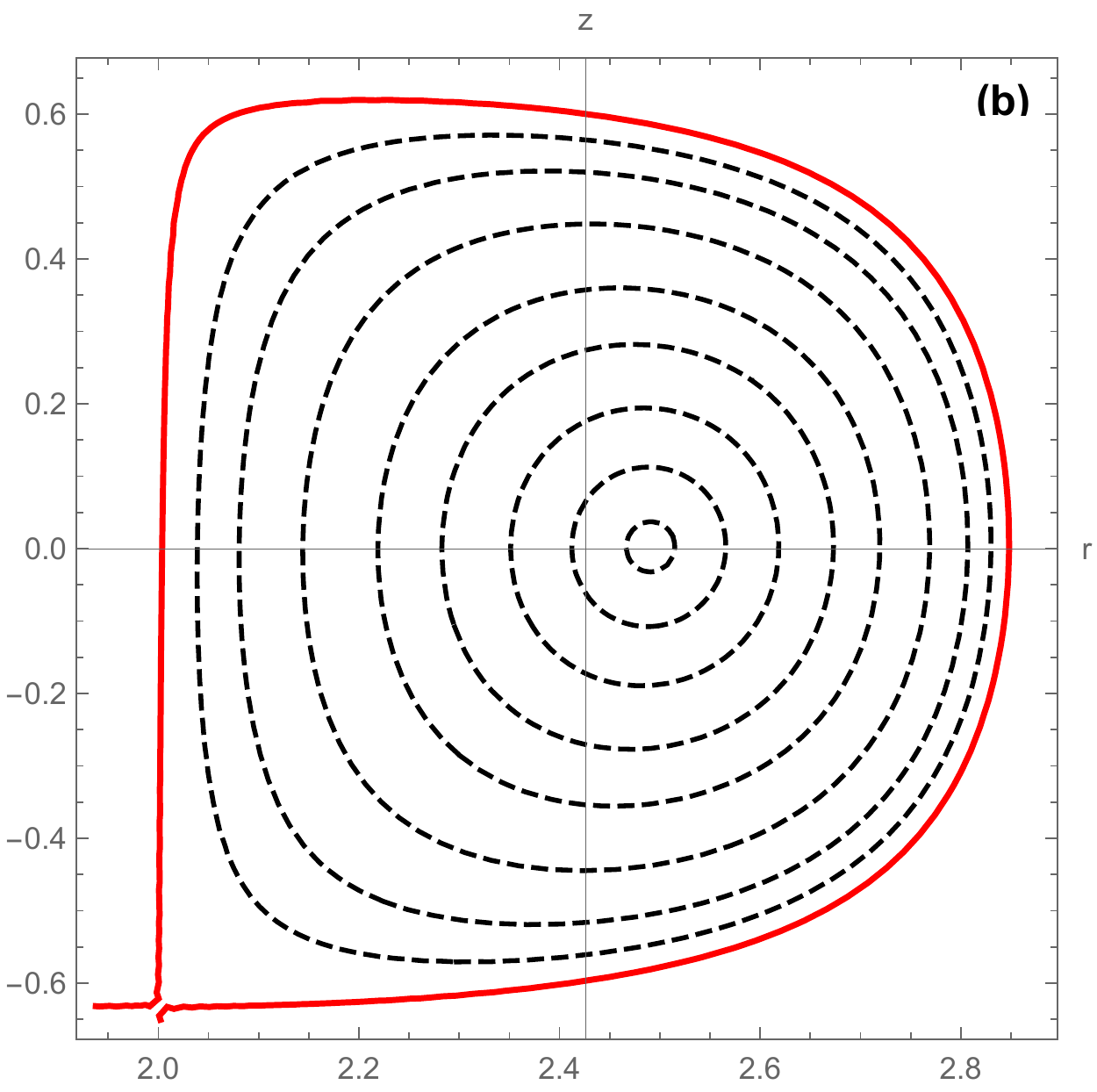} 
\end{center}
\textbf{Fig. 1}. Two sets of closed and nested magnetic surfaces contained in separatrices. The figure on the left (a) corresponds to a compact magnetic configuration with geometric features relevant to the Sustained Spheromak Physics Experiment (SSPX) \cite{SSPX}. The D shape of the boundary, chosen to coincide with the separatrix, is due to the non parallel flow contribution. In the absence of such a flow and given that $s^+=s^-$, the separatrix is rectangular. On the right (b) is depicted a non-compact, up-down asymmetric configuration with inverse aspect ratio $\varepsilon\simeq 0.17$ and elongation $\kappa\simeq 1.5$. Both configurations are constructed using $\tilde{u}_1$ with  $h_1$ $\simeq -0.2$ and $h_1\simeq -0.3$, respectively.
\begin{center}
\includegraphics[scale=0.345]{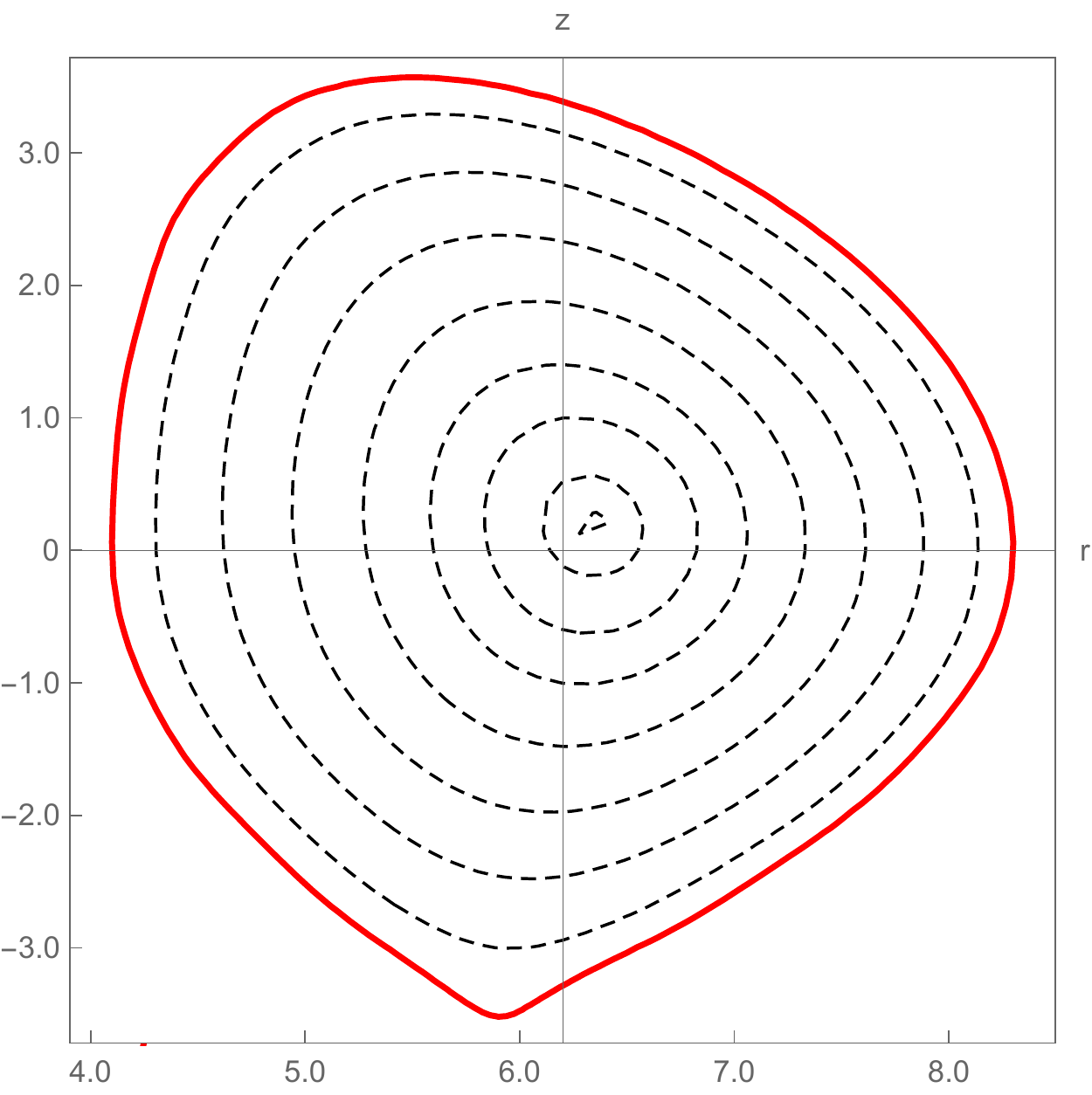} \label{ITER}
\end{center}
\textbf{Fig. 2}. A shaped Tokamak configuration obtained by the level sets of $\tilde{u}_2$. The geometric characteristics are similar to those of the International Thermonuclear Experimental Reactor (ITER) \cite{ITER}.
\end{figure}
\paragraph{Compact toroids} Setting the parameter $\lambda_1$ in Eq. (\ref{sol_lin_1}) equal to zero, one can identify a set of closed and nested magnetic surfaces, which is suitable for the description of compact configurations. In order to bound the plasma into a predetermined region we may use the  condition $\tilde{u}_1(r=R_s,z=0)=\tilde{u}_b$, where $\tilde{u}_b$ is the value of the function $\tilde{u}_1$ on the axis of symmetry and $(R_s,0)$ the outer point of the separatrix which is chosen as the outer closed magnetic surface of this particular set. In Fig. 1(a) is depicted an isolated set of magnetic surfaces which corresponds to a compact-toroid  with $R_s=0.5$ $m$. 
\paragraph{Non compact toroids-Tokamak} Employing $\tilde{u}_1$ we can also identify sets of magnetic surfaces suitable for the description of non compact configurations with various aspect ratios. An isolated set of such surfaces which could describe a Tokamak configuration is depicted in Fig. 1(b). The red-colored closed
line corresponds to the separatrix of the given set. The up-down asymmetry is due to the slightly different values of the parameters $s^+$ and $s^-$. In both cases of compact and non compact configurations, the flow parameter $h_1$ affects the shape of the separatrix; for $h_1=0$ and $s^+=s^-$ the separatrix has rectangular shape while as $|h_1|$ increases the equilibria acquire D-shaped boundaries.
\paragraph{Shaped Tokamak} Using $\tilde{u}_2(r,z)$ and following the shaping method described in \cite{Cerf_Freid} and \cite{Kuir_Throum}, we exploited the arbitrary parameters $\kappa_{1,j},...,\kappa_{4,j}$ with $j=1,...,8$, to construct shaped equilibrium with a cross section relevant to ITER (Fig. 2).
\paragraph{Linear force-free equilibria} Another case of linear equilibrium solutions are those obtained in the limit $g(u)\rightarrow 0$ by solving Eq. (\ref{ff_ode}) with $f(u)=f_1+f_2 u$. These solutions describe Taylor force-free relaxed states. The completely free parameters $a$, $b$ and the linearity of the equilibrium equation allows general solutions expressed by means of arbitrary number of terms:
\begin{eqnarray}
u_{ff}^{\pm}(r,z)=\sum_{j=1}^N  \biggl[ \biggr. \mu_{1,j}cos\left(\sqrt{f_2\left(\beta_j^2+r^2+z^2\pm 2\beta_j z\right)}\right) \nonumber \\
+\mu_{2,j}sin\left(\sqrt{f_2\left(\beta_j^2+r^2+z^2\pm 2\beta_j z\right)}\right)\biggl. \biggr]-\frac{f_1}{f_2}
\end{eqnarray}
where $\beta_j$ and $\mu_{1,j}$, $\mu_{2,j}$ with $j=1,...,N$ are arbitrary parameters. 

\subsection{Nonlinear solutions} 
The reduction methods described previously are now implemented for several nonlinear choices of free functions $f(u)$, $g(u)$, with the reminder that $g(u)=\epsilon f(u)$. Here we give some examples of analytically calculated solutions for several nonlinear choices of free functions. Note that employing Eq. (\ref{w}) we can integrate it for $c \neq 0$ or $c=0$. It may turn out that some solutions which belong to the former case are not reducible to solutions of the latter class just by setting $c=0$, see for example the solutions denoted by the subscript $1$ below. The examples presented are not exhaustive and possibly one may find additional solutions for these particular or additional free functions. 
\paragraph{Quadratic} $f(u)=f_0+f_1u^2$. In this case we observe that Eq. (\ref{dir_red_ode}) takes the form of a differential equation satisfied by the Weierstrass elliptic function $\wp$. The exact solutions are given by:
 \begin{equation}
 u_{u^2}^{\pm}(r,z)=\eta\wp\left(\eta^{-1}\left(\frac{a r^2}{2}\pm bz+\tilde{c}\right);\frac{2f_0\eta}{b^2},c\right) \label{weier}
 \end{equation}
 where $\eta:=6^{1/3}\left(-f_1/b^2\right)^{-1/3}$.
\paragraph{Exponential} $f(u)=f_0e^{nu}$ with $n\in \mathbb{R}$ and $f_0$ an arbitrary constant, we present two classes of solutions:
\begin{eqnarray}
u_{exp,1}^{\pm}(r,z)= \frac{1}{n} ln\left(\frac{b^2 c n} {2f_0}sech^2\left(\frac{n\sqrt{c}}{4}\left(a r^2\pm 2bz+\tilde{c}\right)\right)\right) \label{exp_1} \\
u_{exp,2}^{\pm}(r,z)=-\frac{2}{n}ln\left(-i\frac{\sqrt{2f_0n}}{2b}\left(\frac{ar^2}{2}\pm bz+\tilde{c}\right)\right) \label{exp_2}
\end{eqnarray}
\paragraph{Integer exponent} $f(u)=f_0u^n$ with $n\in \mathbb{Z}\setminus\{1\}$. In this case the solutions (\ref{pow}) are obtained explicitly from (\ref{w}) with $c=0$.
\begin{eqnarray}
u_{u^n}^{\pm}(r,z)= 
2^{\frac{2}{n-1}}\left[\frac{i(1-n)}{b}\sqrt{\frac{2f_0}{n+1}}\left(\frac{ar^2}{2}\pm bz\right)+\tilde{c}\right]^{\frac{2}{1-n}} \label{pow}
\end{eqnarray}
\paragraph{Sinusoidal} $f(u)=f_0sin(nu)$ with $n\in \mathbb{R}$. We find a first class of exact solutions given by (\ref{sin_1}), where $am$ is the Jacobi amplitude function and a second class fiven by (\ref{sin_2}). 
\begin{eqnarray}
u_{sin,1}^{\pm}(r,z)=(\pm) \frac{2}{n} am\left(\frac{\sqrt{nc}}{2b}\left(\frac{ar^2}{2}\pm bz+\tilde{c}\right)\bigg| \frac{4f_0}{c}\right) \label{sin_1} \\
u_{sin,2}^{\pm}(r,z)=\nonumber \\
(\pm)\frac{4}{n} arccot\left(exp\left(i\frac{\sqrt{f_0n}}{b}\left(\frac{ar^2}{2}\pm bz+\tilde{c}\right)\right)\right)\label{sin_2} 
\end{eqnarray}
\paragraph{Force-free} $g(u)=0$, $f(u)=f_0e^{n u}$ with $n\in \mathbb{R}$. Solving Eq. (\ref{ff_ode}) we obtain the following solutions which describe nonlinear force-free equilibria,
\begin{eqnarray}
u_{exp,ff}^{\pm}(r,z)= \nonumber \\ \frac{1}{n} ln\left(cn^2 sech^2\left(\frac{\sqrt{2n^3cf_0[4a(ar^2+az^2\pm bz)+b^2]}+\tilde{c}}{4a}\right)\right)
\label{ff}
\end{eqnarray}
In Fig. 3 are depicted the level sets of solutions (\ref{weier}) and (\ref{ff}). Both of the depicted equilibria possess current sheet formations.  Also, by changing the parametric values one can construct respective configurations without current sheets. The topology of the magnetic surfaces of the finite pressure-gradient equilibria displays a parabolic morphology; an example is shown in Fig. 3(a). This characteristic feature comes from the special form of the reduction ansatz. 
\begin{figure}
\begin{center}
\includegraphics[scale=0.34]{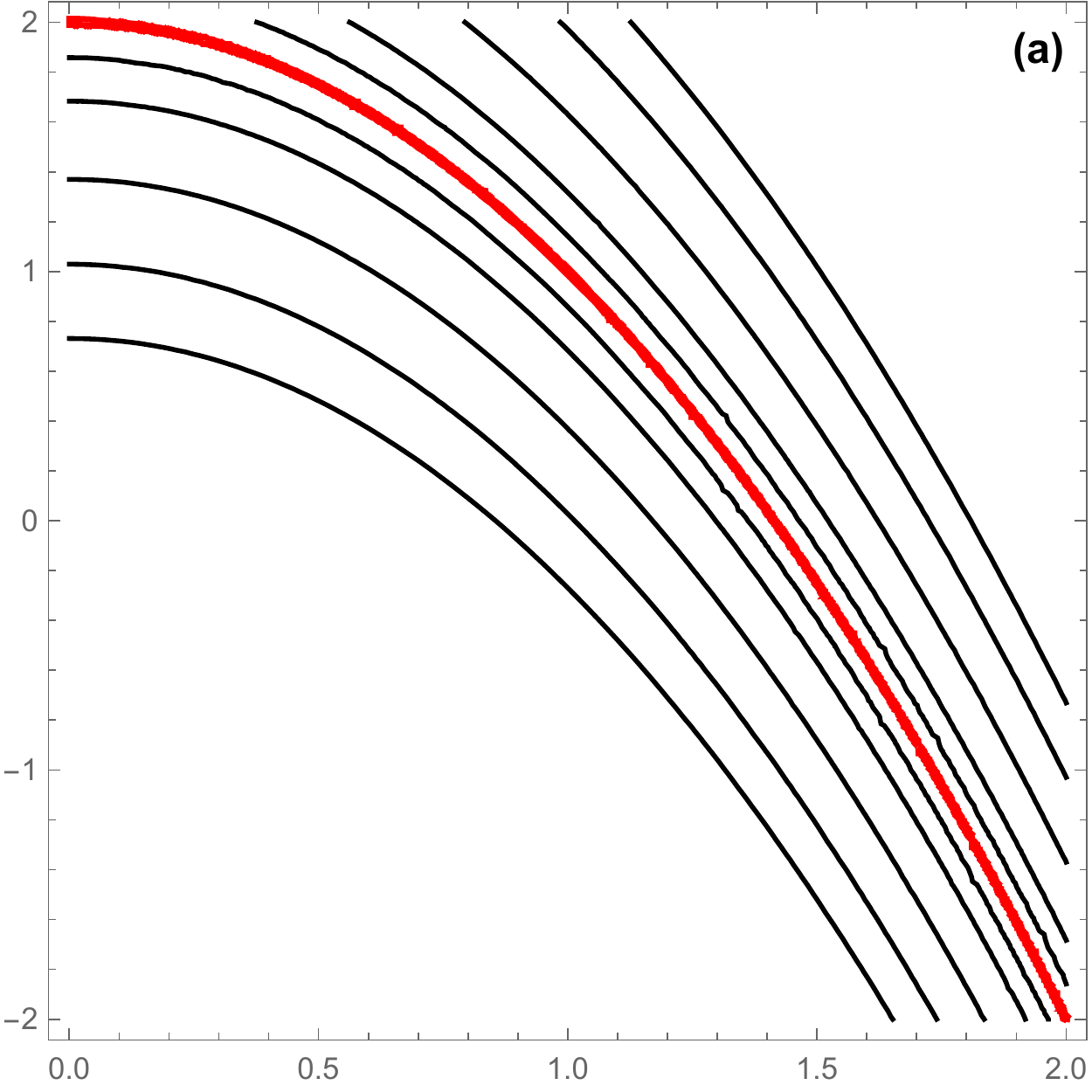} \includegraphics[scale=0.35]{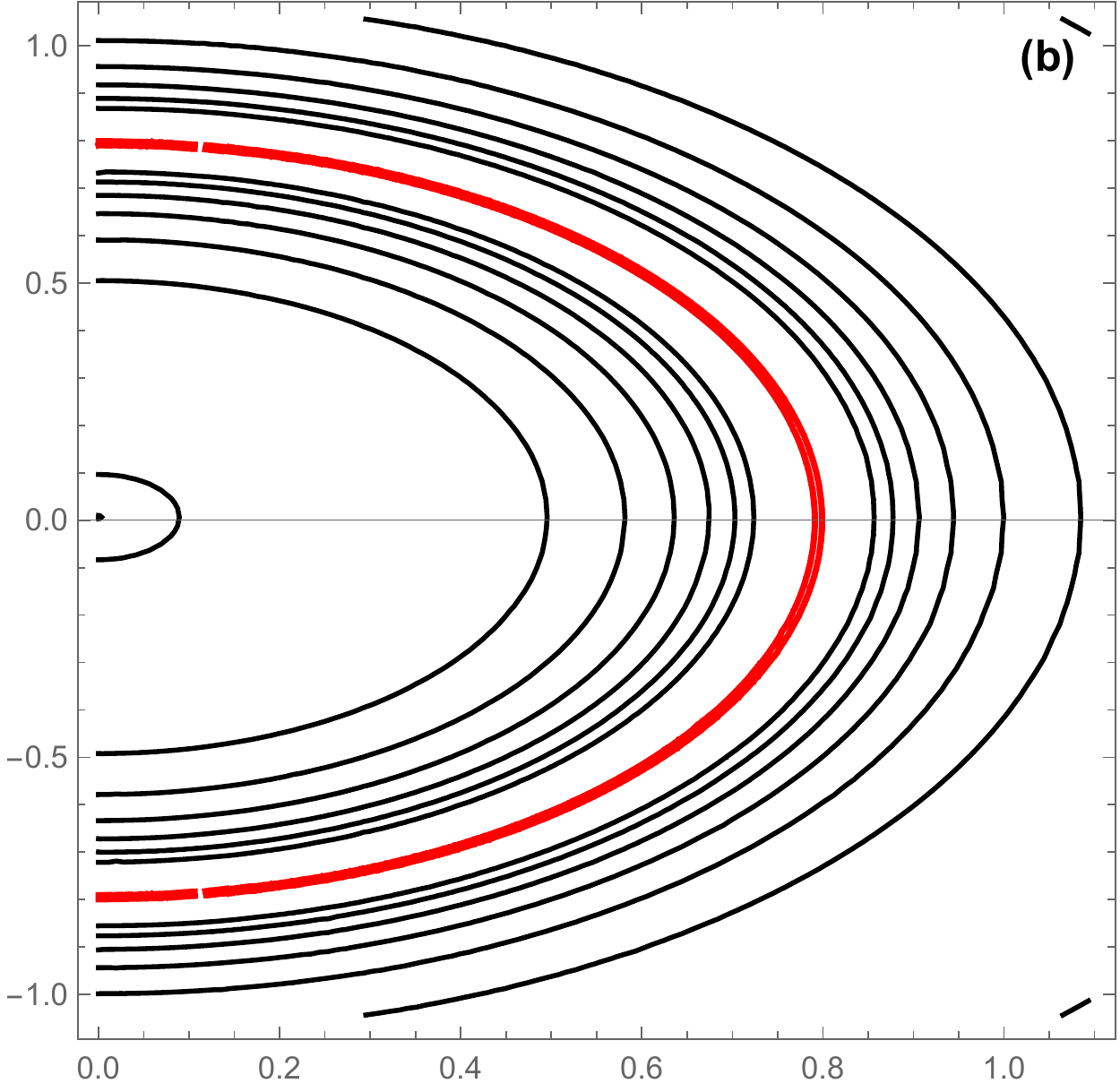} \label{para}
\end{center}
\textbf{Fig. 3}. The level sets of the functions (a) $u^{+}_{u^2}$ and(b) $u_{exp,ff}^{0}$ (i.e. $b=0$) on the plane $(r,z)$. The red thick contours corresponding to level sets where $u$ isn't differentiable represent current sheets. Configurations without current sheets are also possible by changing  the parametric values.
\end{figure}
\section{Discussion}
We remark here that due to the nonlinearity of the equilibrium equation the linear combinations of different solutions which possibly could result in closed magnetic surfaces are ruled out. However it is probable that the aforementioned similarity reduction methods can be employed using alternative reduction ansatzes leading to solutions with desirable characteristics. Particularly in the framework of the group foliation method, additional symmetries of the GS equation \cite{Cic_Peg_1} can possibly be exploited in order to obtain alternative reductions. This possibility though, is restricted
to just few particular choices of free functions in contrast with the
reductions obtained by the translational symmetry, which is an intrinsic property of the GS equation, regardless the choice of free
functions.

Hypothetically a nonlinear superposition principle could be used in order to combine nonlinear solutions. However it turns out that in the case of GS equation the classical method for establishing such superposition principles \cite{Jones_Ames,Levin}, where the combination of different solutions is realized through a so-called, reduced connecting function, is of no practical use. Thus it is an open question whether any nonlinear or pseudo-linear superposition principles can be applied or established for the GS equation. If this will be proved possible, then the nonlinear solutions obtained here may give us new insights in the analytical study of axisymmetric MHD equilibria.
 \section*{Acknowledgements}
The authors would like to thank Dimitris Tsoubelis, Henri Tasso,
George Poulipoulis and the anonymous reviewers for useful comments.

This work has been carried out within the framework of the EUROfusion Consortium and has received funding from  (a) the National Programme for the Controlled Thermonuclear Fusion, Hellenic Republic, (b) Euratom research and training programme 2014-2018 under grant agreement No 633053. The views and opinions expressed herein do not necessarily reflect those of the European Commission.
 
 \section*{References}


\end{document}